\documentclass{article}
\usepackage{amsfonts}
\usepackage{amssymb}
\usepackage{amsmath}
\usepackage{amsthm}
\usepackage[english]{babel}
\begin{document}

\title{\bf The geometrical meaning\\ of the Weitzenb{\" o}ck connection}

\author{{\bf Alexey Golovnev}\\
{\small {\it Centre for Theoretical Physics, The British University in Egypt,}}\\
{\small \it El Sherouk City, Cairo 11837, Egypt}\\
{\small agolovnev@yandex.ru}}
\date{}

\maketitle

\begin{abstract}

In the current literature, there are many discussions about the local Lorentz invariance of modified teleparallel gravity. This symmetry is obviously violated in the classical "pure tetrad" formulation of the theory, while it gets restored in the "fully covariant" approach. My claim is that, despite many heated discussions, the two formulations are just equivalent. And the purpose of this note is to argue that the local Lorentz invariance is not natural for the modified teleparallel theories at all, making the pure tetrad approach more fundamentally justified.

\end{abstract}

\section{Introduction}

I would not go into many details here, however we all know that there are lots of reasons to try modifying gravity, with the common backgrounds of this wish varying from problems of quantum gravity to observational tensions in cosmology. And to our sorrow or to our fun, we also know how difficult it is to non-trivially modify the structure of General Relativity (GR) without immediately running into much more severe troubles, often of mere self-consistency, let alone any hope for a vague hint of well-posedness and stability.

Despite all these complications, the more we play with many simple options of adding new degrees of freedom, the clearer it seems that one should also examine radical changes of geometry, with some of the ideas dating back to Einstein himself. Among various other lines of research, here comes teleparallel gravity \cite{Pereira} and its modifications \cite{Hayashi,Rafael}. There are many reasons to believe that numerous modifications of this sort have little chance of being viable \cite{problems}. However, it would be very interesting and useful for our understanding of gravity to invest some good effort into realising how it works and what are the sources of troubles. In particular, this text of mine got its inspiration from a very nice conference held in Tartu, Estonia: Metric-Affine Frameworks for Gravity 2022.

My main topic here is about the issue of local Lorentz invariance in (metric-compatible) modified teleparallel gravities. There are many discussions on the covariant approach \cite{KraSar,GKS} to (modified) teleparallel theories. The conclusions range all the way from total denial \cite{Maluf} through an undecided stance \cite{Maria} to taking it as the only consistent version of the theory \cite{Martin}. In this paper, I would like to back the middle position. On one hand, I will argue that the pure tetrad approach is {\it the} natural geometric approach to teleparallel descriptions. On the other hand, the covariant rewriting has nothing wrong in itself either.

Let me now very briefly review the teleparallel framework, mostly for fixing my notations. In particular, I prefer to put the derivative-related index in the connection coefficient to the left: 
$$\bigtriangledown_{\mu} A^{\alpha}=\partial_{\mu} A^{\alpha} + \Gamma^{\alpha}_{\mu\nu} A^{\nu}.$$ On top of the usual curvature tensor $R^{\alpha}_{\hphantom{\alpha}\beta\mu\nu}$, an arbitrary connection $\Gamma^{\alpha}_{\mu\nu}$ on a metric manifold can be characterised by two more tensors: torsion 
$$T^{\alpha}_{\hphantom{\alpha}\mu\nu}=\Gamma^{\alpha}_{\mu\nu} - \Gamma^{\alpha}_{\nu\mu} $$ 
and non-metricity 
$$Q_{\alpha\mu\nu}=\bigtriangledown_{\alpha}g_{\mu\nu}.$$ 
Teleparallel theories do use metric-affine constructions with zero curvature, $R^{\alpha}_{\hphantom{\alpha}\beta\mu\nu}=0$, and I will be discussing its old and relatively well-studied variant with no non-metricity either,  $Q_{\alpha\mu\nu}=0$, which means gravity description in terms of torsion. 

With simple necessary modifications, one can repeat the textbook derivation of the Levi-Civita connection ${\mathop \Gamma\limits^{(0)}}{}^{\alpha}_{\mu\nu}$ and find an arbitrary connection in terms of the Levi-Civita one and contributions from torsion and non-metricity. Assuming that $Q_{\alpha\mu\nu}=0$, though not yet restricting the curvature, we get
\begin{equation}
\label{connvar}
\Gamma^{\alpha}_{\mu\nu}={\mathop \Gamma\limits^{(0)}}{}^{\alpha}_{\mu\nu} + K^{\alpha}_{\hphantom{\alpha}\mu\nu}
= {\mathop \Gamma\limits^{(0)}}{}^{\alpha}_{\mu\nu} + \frac12 \left(T^{\alpha}_{\hphantom{\alpha}\mu\nu} + T^{\hphantom{1}\alpha}_{\mu\hphantom{\alpha}\nu} + T^{\hphantom{1}\alpha}_{\nu\hphantom{\alpha}\mu}  \right)
\end{equation}
with an antisymmetric in the lateral indices contortion tensor $K_{\alpha\mu\nu}=-K_{\nu\mu\alpha}$.

Substituting this beautiful result (\ref{connvar}) into the standard definition of the curvature tensor and making the necessary contractions, it's easy to see that the scalar curvatures of the two connections are related to each other in a very simple way:
\begin{equation}
\label{basrel}
R={\mathop R\limits^{(0)}} + 2 {\mathop \bigtriangledown\limits^{(0)}}_{\mu} T^{\mu} + \mathbb T
\end{equation}
where the superscript of ${}^{(0)}$ denotes the quantities calculated with the Levi-Civita connection, $T^{\mu}$ is the torsion vector
\begin{equation}
\label{tvec}
T_{\mu}= T^{\alpha}_{\hphantom{\alpha}\mu\alpha},
\end{equation}
and the very important quantity is the torsion scalar
\begin{equation}
\label{tscal}
{\mathbb T}= \frac12 S^{\alpha\mu\nu} T_{\alpha\mu\nu}  = \frac12 \left(\vphantom{\int} K^{\mu\alpha\nu} + g^{\alpha\mu}T^{\nu} - g^{\alpha\nu}T^{\mu} \right) T_{\alpha\mu\nu}
\end{equation}
with $S^{\alpha\mu\nu}=-S^{\alpha\nu\mu}$ often called superpotential. 

Given the formula (\ref{basrel}), we see that if a curvatureless connection, therefore with $R=0$, is found and is compatible with the arbitrary metric used, then
\begin{equation}
\label{R-T}
-{\mathop R\limits^{(0)}} = 2 {\mathop \bigtriangledown\limits^{(0)}}_{\mu} T^{\mu} + \mathbb T,
\end{equation}
and therefore an action $S=\int d^4 x \sqrt{-g} {\mathbb T}$ gives the same equations as $S=-\int d^4 x \sqrt{-g} {\mathop R\limits^{(0)}}$, and is called the Teleparallel Equivalent of General Relativity (TEGR). At the same time, almost every modification of it does bring us far away from other familiar models of gravity.

Finally, let me mention that the standard way of working in teleparallel gravity models is in terms of tetrads. A tetrad at a given point of a manifold is a basis of its tangent space which means a full set of linearly-independent vectors. Obviously, every vector has its dual element in the cotangent space, and therefore a co-tetrad also exists, i.e. a basis in the cotangent space. Another name for a tetrad is the German word "Vierbein", which can also be nicely transformed to arbitrary dimension by changing one single letter: "Vielbein". However, most of the modern teleparallel literature prefers to use the word "tetrad", even when beyond 4D in clear contradiction to its Greek root.

\section{The meaning of teleparallel geometry}

We call geometry teleparallel if the results of a parallel transport do not depend on a smooth path taken. It means that, given a (co-)tetrad $e^a_{\mu}$ at a given point, one can parallelly transport it to every other point on the manifold and get a uniquely defined covariantly constant field $e^a_{\mu}(x)$:
\begin{equation}
\label{teldef}
\bigtriangledown_{\nu}e^a_{\mu}=0.
\end{equation}
Note that it's not a soldering form here. For that we would take the zero covariant derivative as a definition of the spin-connection. However, I take the index ${}^a$ as simply a number of a 1-form. And I assume that those are four separate 1-forms which define the notion of parallel transport on the manifold, by being covariantly constant on it.

Of course, at every point this basis $e^a_{\mu}$ of the cotangent space is dual to some basis $e^{\mu}_a$ of the tangent space, and we could have gone in terms of a tetrad instead of co-tetrad: $\bigtriangledown_{\nu}e_a^{\mu}=0$, with the index ${}_a$ again just numbering the vectors. Moreover, following big parts of the teleparallel literature, I will allow myself to omit the prefix "co-", even when talking about the cotangent space basis.

By definition, we take the co-tetrad of the formula (\ref{teldef}) as a basis of covariantly conserved 1-forms:
$$0=\bigtriangledown_{\mu}e^a_{\nu}=\partial_{\mu}e^a_{\nu}-\Gamma^{\alpha}_{\mu\nu}e^a_{\alpha}$$
which immediately implies the Weitzenb{\" o}ck relation
\begin{equation}
\label{condef}
\Gamma^{\alpha}_{\mu\nu}=e^{\alpha}_a \partial_{\mu}e^a_{\nu}
\end{equation}
with the tetrad, i.e. a basis of vectors, $e^{\alpha}_a$ being the matrix inverse of the co-tetrad.

Globally, such a covariantly conserved field might not exist at all, due to purely topological obstructions. Indeed, already Poincar{\'e} and Brouwer realised that it was impossible to comb a hedgehog. However, on a given coordinate chart, it's always possible to find such a connection. In a sense, every teleparallel geometry in a given coordinate chart does have such a tetrad field which is covariantly constant (\ref{teldef}), or to put it more precisely, an equivalence class of such fields. We might have chosen another 1-form basis at the initial point, ${\tilde e}^a_{\mu}=L^a_b e^b_{\mu}$ with an arbitrary non-degenerate matrix $L^a_b$. Then we can define another covariantly constant tetrad field,  ${\tilde e}^a_{\mu}(x)=L^a_b\cdot e^b_{\mu}(x)$, equivalent to the initial one $e^a_{\mu}(x)$ in the sense of defining the same geometry, ${\tilde\Gamma}^{\alpha}_{\mu\nu}=\Gamma^{\alpha}_{\mu\nu}$. 

In other words, the connection (\ref{condef}) is invariant under global general linear transformations of its defining covariantly conserved tetrad. In particular, at any point of the space(time), we can represent an arbitrary vector, or an arbitrary 1-form, as a linear combination of the defining tetrad, or co-tetrad, basis elements $A_{\mu}=A_a e^a_{\mu}$, and then its parallel transport from $x_0$ to $x_1$ is uniquely defined by $e^a_{\mu}(x_0)\longrightarrow e^a_{\mu}(x_1)$ while keeping the coefficients $A_a$ intact (what would be called the frame of zero spin-connection by the people of covariant approach). Therefore, it is obvious that the parallel transport rule given by the connection (\ref{condef}) cannot have any non-trivial holonomy, even without explicitly calculating its vanishing curvature tensor.

At the same time, the connection (\ref{condef}) is {\it not} invariant under local linear transformations of the tetrad field, not even with respect to any particular subgroups of those, like conformal transformations or Lorentz transformations. What I would like to specifically stress is that, due to its very geometric meaning,
\begin{center}
{\it the teleparallel connection should not be invariant\\ under local transformations of its defining tetrad.}
\end{center}
Indeed, the geometrical picture assumes that the tetrad $e^a_{\mu}$ should not be taken as a rank-two tensor with a choice of having two different types of indices. We treat it instead as a set of 1-forms, so that ${}^a$ is not a tensorial index at all.

To put it yet another way, we are not free to choose this tetrad (except a global transformation). On the contrary, it is a very important geometric quantity which is defined by the very geometry at hand. There are many possible teleparallel geometries on a given topological space, and the choice is made by presenting a basis of vectors at every point which are considered as parallel transports of each other. There is no way to have invariance under local linear transformations of this basis. Assuming such an invariance would mean an absolute voluntarism in the notion of parallel transport whose non-uniqueness would not then be restricted to holonomies only, it would rather be a decision to call an arbitrary smooth transport parallel.

\section{The case of no non-metricity}

The discussion above was about an arbitrary geometry of teleparrallel type, i.e. with zero curvature. Let us now demand that there is no non-metricity either. It means that 
$$Q_{\alpha\mu\nu}=\bigtriangledown_{\alpha} g_{\mu\nu}=0,$$ 
and therefore the norms and the scalar products of covariantly constant vectors must be constant, too, and analogously for the covariantly constant 1-forms,
$$\partial_{\mu} M^{ab}=0\quad \mathrm{where} \quad M^{ab}\equiv g^{\mu\nu} e^a_{\mu} e^b_{\nu} \quad 
\mathrm{with\ its\ matrix\ inverse}\quad M_{ab}\equiv g_{\mu\nu} e^{\mu}_a e^{\nu}_b.$$

We can easily invert this relation. Therefore, if at the initial point we have chosen a (co-)tetrad with the matrix $M^{ab}$ of its scalar products, then, upon its parallel transport all over the manifold, the metric can be uniquely expressed everywhere in terms of the tetrad field as
\begin{equation}
\label{genmet}
g_{\mu\nu}(x) = M_{ab} e^a_{\mu}(x) e^b_{\nu}(x).
\end{equation}
The usual assumption of metric teleparallel theories is that the tetrad is orthonormal, that is $M_{ab}=\eta_{ab}$.

In other words, our usual approach to (metric) teleparallelism is to simply postulate that 
\begin{equation}
\label{usemet}
g_{\mu\nu} = \eta_{ab} e^a_{\mu} e^b_{\nu}
\end{equation}
everywhere. At the same time, the variational principle goes in terms of a fully arbitrary (though non-degenerate) tetrad, while the metric (\ref{usemet}) is {\it defined} in terms of the tetrad so that every possible tetrad is by definition orthonormal.

Even in this case, any given metric geometry still has an infinite freedom of choosing its corresponding teleparallel geometry. Indeed, the definition (\ref{usemet}) is invariant under local Lorentz rotations, by the very meaning of the Lorentz group; however, precisely as in the general case above, the connection (\ref{condef}) is invariant only under the global version of those. Once more, the meaning of defining a teleparallel geometry is in choosing a preferred tetrad which is a basis of covariantly constant vectors. Taking the metric-compatible  teleparallel approach means that the choice of the tetrad also restricts the metric by the formula (\ref{usemet}), or more generally by the formula (\ref{genmet}) with a constant non-degenerate matrix $M_{ab}$. However, local rotations of the defining tetrad, even though keeping the metric intact, do change the teleparallel geometry which is not invariant under such transformations.

And precisely as in the general case of the previous Section, let me state it once more:
\begin{center}
{\it the Weitzenb{\" o}ck connection should not be invariant\\ under local Lorentz transformations of its defining tetrad.}
\end{center}
The defining tetrad of a metric-compatible teleparallel geometry is a covariantly-constant orthonormal basis of vectors, or 1-forms, and by no means can it be invariant under local Lorentz rotations, unless we want to call every smooth transportation parallel when it keeps scalar products intact.

All in all, a defining tetrad of the teleparallel geometry in the pure tetrad approach is a preferred frame indeed, however it is an {\it objectively} preferred frame. This preference comes from the very geometry at hand, and the equations of motion determine which frame can be preferred and which cannot. It is an important message we must learn from the modified teleparallel models. Their tetrad is a dynamical variable, it has no relation to the freedom of observers' choice. Sometimes people talk about "good" and "bad" tetrads \cite{goodbad}. It would be a meaningful language if it was about our free choice of measuring devices  which could of course be helpful ("good") or harmful ("bad"). However, in this case, it is simply about our successful or failed Ans{\"a}tze for solving equations.

\section{Theories of $f(T)$ type with non-orthonormal tetrads}

One can of course start with a non-orthonormal tetrad. That would mean choosing a non-degenerate matrix $M_{ab}$ instead of the Minkowski one; once and for all, for it must then be the same all over the spacetime manifold. For example, null tetrads were previously used \cite{null,null2}. This is actually a fully legitimate choice, too. 

Indeed, let's take any particular point on a spacetime manifold with a tetrad $e^a_{\mu}$ of a particular matrix $M_{ab}$ of its scalar products (\ref{genmet}). If the manifold has a metric-compatible teleparallel structure on it, we then get a full covariantly-constant tetrad field $e^a_{\mu}(x)$, everywhere with its same matrix $M_{ab}$. One can take a normalising transformation of variables
$$e^a_{\mu} \longrightarrow  {\tilde e}^a_{\mu} = N^a_b e^b_{\mu}$$
with $\partial_{\mu} N^a_b =0$ and such that
$$g_{\mu\nu}=M_{ab} e^a_{\mu} e^b_{\nu}= \eta_{ab} {\tilde e}^a_{\mu} {\tilde e}^b_{\nu}.$$
Obviously, it is enough to find a solution of an equation $M_{cd}=N^a_c \eta_{ab} N^b_d$ for that. In the ${\tilde e}^a_{\mu}$ variables, I would have the usual equations of motion, as in the standard teleparallel theories with orthonormal tetrads. At the same time, the equations for $e^a_{\mu}$ are of course equivalent to those for ${\tilde e}^a_{\mu}$ since $\frac{\delta S}{\delta e^a_{\mu}}=N^b_a \cdot \frac{\delta S}{\delta {\tilde e}^b_{\mu}}$.

It is instructive to see how it works in the simplest case of $f(T)$ theories. And I also use this occasion for presenting a nice and covariant way of deriving the equations \cite{GK,issues}. When we have an action
$$S[e^a_{\mu}(x)]=\int d^4 x \sqrt{-g} f({\mathbb T}) \quad \mathrm{where} \quad T^{\alpha}_{\hphantom{\alpha}\mu\nu}=  e^{\alpha}_a \left(\partial_{\mu}e^a_{\nu} -  \partial_{\nu}e^a_{\mu}\right) \quad \mathrm{and} \quad g_{\mu\nu}(x) = M_{ab} e^a_{\mu}(x) e^b_{\nu}(x),$$
with the usual definitions of the torsion scalar (\ref{tscal}) and the torsion vector (\ref{tvec}) and the relation (\ref{R-T}) guaranteed, a very nice way of doing the variations is
\begin{equation}
\label{thevar}
\delta S= \int d^4 x \left( f({\mathbb T}) \delta\sqrt{-g} + \sqrt{-g} f^{\prime}({\mathbb T}) \delta{\mathbb T} \right)= \int d^4 x \sqrt{-g} \left(f\cdot \frac{\delta\sqrt{-g}}{\sqrt{-g}} - f^{\prime}\cdot \delta {\mathop R\limits^{(0)}} -  2f^{\prime}\cdot \delta ({\mathop \bigtriangledown\limits^{(0)}}{}^{\mu}T_{\mu})  \right).
\end{equation}

For the most part of calculations, it is enough to know that
$$\delta g_{\mu\nu}=M_{ab}\left( e^a_{\mu} \delta e^b_{\nu} +  e^b_{\nu} \delta e^a_{\mu} \right),$$
and to use the well-known variations of the Riemann tensor and the Levi-Civita connection. Then the first and the second terms in the variation (\ref{thevar}) are
$$f\cdot \frac{\delta\sqrt{-g}}{\sqrt{-g}}=\frac12 f g^{\mu\nu} \delta g_{\mu\nu}=\frac12 f g^{\mu\nu}\cdot M_{ab}\left( e^a_{\mu} \delta e^b_{\nu} +  e^b_{\nu} \delta e^a_{\mu} \right) = f g^{\mu\nu}\cdot M_{ab}  e^b_{\nu} \delta e^a_{\mu}$$
and
\begin{multline*}
- f^{\prime}\cdot \delta {\mathop R\limits^{(0)}}= -f^{\prime} \left(-{\mathop R\limits^{(0)}}{}^{\mu\nu}\delta g_{\mu\nu} + g^{\mu\nu} \left( {\mathop \bigtriangledown\limits^{(0)}}{}_{\alpha}\delta{\mathop \Gamma\limits^{(0)}}{}^{\alpha}_{\mu\nu} -  {\mathop \bigtriangledown\limits^{(0)}}{}_{\nu}\delta{\mathop \Gamma\limits^{(0)}}{}^{\alpha}_{\alpha\mu} \right)\right) = f^{\prime} \left({\mathop R\limits^{(0)}}{}^{\mu\nu}- {\mathop \bigtriangledown\limits^{(0)}}{}^{\mu} {\mathop \bigtriangledown\limits^{(0)}}{}^{\nu} + g^{\mu\nu}{\mathop \square\limits^{(0)}}\right)\delta g_{\mu\nu}\\
\longrightarrow 2 \left( \left({\mathop R\limits^{(0)}}{}^{\mu\nu}- {\mathop \bigtriangledown\limits^{(0)}}{}^{\mu} {\mathop \bigtriangledown\limits^{(0)}}{}^{\nu} + g^{\mu\nu}{\mathop \square\limits^{(0)}}\right) f^{\prime} \right) \cdot M_{ab}  e^b_{\nu} \delta e^a_{\mu}
\end{multline*}
respectively.  In the last line, after integration by parts, we have used the fact that the two Levi-Civita covariant derivatives commute with each other when acting on a scalar, and therefore it is not necessary to explicitly symmetrise the matrix $ M_{ab}  e^b_{\nu} \delta e^a_{\mu}$ coming from $\delta g_{\mu\nu}$.

The last term in the variation (\ref{thevar}) can be presented as
\begin{multline*}
-2f^{\prime}\cdot \delta ({\mathop \bigtriangledown\limits^{(0)}}{}^{\mu}T_{\mu}) = -2f^{\prime}\cdot \delta\left(\frac{1}{\sqrt{-g}}\partial_{\mu} \left(\sqrt{-g} g^{\mu\nu} T_{\nu} \right) \right)\\
= -2f^{\prime} \left({\mathop \bigtriangledown\limits^{(0)}}{}^{\mu}\delta T_{\mu} + {\mathop \bigtriangledown\limits^{(0)}}{}_{\alpha}\left( \left(\frac12 g^{\alpha\beta} g^{\mu\nu}- g^{\alpha\mu}g^{\beta\nu}\right)T_{\beta} \delta g_{\mu\nu} \right)- \frac12 ({\mathop \bigtriangledown\limits^{(0)}}{}^{\alpha}T_{\alpha}) \cdot g^{\mu\nu}\delta g_{\mu\nu}\right).
\end{multline*}
The only non-trivial piece is $\delta T_{\mu}$, and it is the only source of the antisymmetric part in the equations of motion. It can be very conveniently found by observing that
\begin{equation}
\label{telconnvar}
\delta\Gamma^{\alpha}_{\mu\nu}=e_a^{\alpha}\partial_{\mu}\delta e^a_{\nu}-(\partial_{\mu}e^a_{\nu}) e_a^{\beta}e^{\alpha}_c \delta e^c_{\beta}=e_c^{\alpha}\partial_{\mu}\delta e^c_{\nu} - e^{\alpha}_c \Gamma^{\beta}_{\mu\nu}\delta e^c_{\beta} = e_a^{\alpha}\bigtriangledown_{\mu}\delta e^a_{\nu}=  \bigtriangledown_{\mu}\left( e_a^{\alpha} \delta e^a_{\nu}\right)
\end{equation}
which immediately yields
$$\delta T_{\mu}=\partial_{\mu}\left( e_a^{\alpha} \delta e^a_{\alpha}\right)-  \bigtriangledown_{\alpha}\left( e_a^{\alpha} \delta e^a_{\mu}\right)=
\partial_{\mu}\left( e_a^{\alpha} \delta e^a_{\alpha}\right)-  {\mathop \bigtriangledown\limits^{(0)}}_{\alpha}\left( e_a^{\alpha} \delta e^a_{\mu}\right) - K^{\alpha}_{\hphantom{\alpha}\alpha\beta} e_a^{\beta} \delta e^a_{\mu} + K^{\beta}_{\hphantom{\beta}\alpha\mu} e_a^{\alpha} \delta e^a_{\beta}.$$
Using the fact that $K^{\alpha}_{\hphantom{\alpha}\alpha\beta}=-T_{\beta}$ and obvious transformations such as $T_{\beta} e^{\beta}_a=T^{\alpha} g_{\alpha\beta} e^{\beta}_a= T^{\alpha} M_{ab} e^b_{\alpha}$, altogether we get
\begin{multline*}
-2f^{\prime}\cdot \delta ({\mathop \bigtriangledown\limits^{(0)}}{}^{\mu}T_{\mu}) \longrightarrow 2\cdot \left( \left(- g^{\mu\nu}{\mathop \square\limits^{(0)}} +  {\mathop \bigtriangledown\limits^{(0)}}{}^{\mu} {\mathop \bigtriangledown\limits^{(0)}}{}^{\nu} \right)f^{\prime} +  (\partial_{\alpha} f^{\prime})\cdot \left(T^{\nu}g^{\alpha\mu} +K^{\mu\nu\alpha}\right)  \right.\\
+\left. (\partial_{\alpha} f^{\prime})\cdot \left(T^{\alpha}g^{\mu\nu}-T^{\mu}g^{\alpha\nu}-T^{\nu}g^{\alpha\mu} \right) +  ({\mathop \bigtriangledown\limits^{(0)}}{}^{\alpha}T_{\alpha}) \cdot f^{\prime}g^{\mu\nu} \right) \cdot M_{ab}  e^b_{\nu} \delta e^a_{\mu}
\end{multline*}
with the first line coming from ${\mathop \bigtriangledown\limits^{(0)}}{}^{\mu}\delta T_{\mu}$ and the second one from everything else.

Now, summing it all, the variation (\ref{thevar}) takes the form of
$$\delta S =2 \int d^4 x \sqrt{-g} \left(f^{\prime} {\mathop R\limits^{(0)}}{}^{\mu\nu} +\left(\frac12 f + f^{\prime}{\mathop \bigtriangledown\limits^{(0)}}{}^{\alpha}T_{\alpha} \right) g^{\mu\nu} + (\partial_{\alpha} f^{\prime})\cdot \left(K^{\mu\nu\alpha} + T^{\alpha}g^{\mu\nu}-T^{\mu}g^{\alpha\nu} \right) \right)M_{ab}  e^b_{\nu} \delta e^a_{\mu},$$
with all the higher derivative terms successfully cancelled out. Given the non-degeneracy of the matrix $M_{ab}  e^b_{\nu}$, and using the relation (\ref{R-T}) and the definition of the superpotential tensor (\ref{tscal}), we get the usual equation \cite{coveq}
\begin{equation}
\label{eqomo}
f^{\prime} {\mathop G\limits^{(0)}}{}^{\mu\nu} +\frac12 \left( f - f^{\prime}{\mathbb T} \right) g^{\mu\nu} + f^{\prime\prime} S^{\nu\mu\alpha}\partial_{\alpha}\mathbb T=0.
\end{equation}
On top of reproducing the known equation (\ref{eqomo}), note also that the $f^{\prime\prime}$-term comes from variations of the derivative of the tetrad inside the torsion tensor, therefore getting it proportional to precisely the superpotential from the definition of the torsion scalar (\ref{tscal}) is a nice cross-check, too.

As we see, equations of modified teleparallel gravity are covariant under diffeomorphisms. Unfortunately, for a long time even the simplest models such as $f(T)$ gravity were used to be written in strikingly non-covariant ways. To my knowledge, the covariant shape of equations for $f(T)$ first appeared in the Ref. \cite{coveq} being derived from the standard non-covariant one. The simple derivation presented here and in Refs. \cite{GK, issues} is also applicable only to $f(T)$ and $f(T,B)$ models, with $B$ staying for the boundary term $B=2{\mathop \bigtriangledown\limits^{(0)}}{}_{\mu}T^{\mu} $. However, the basic observation about  variation of the teleparallel connection (\ref{telconnvar})   can perfectly be used for any modified telaparallel model including New GR. 

\section{Using anholonomic bases}

Having given all the preferred frame explanations above, let me say it again that I am not against using the "Lorentz-covariant" language when it is convenient. Of course, there is nothing wrong about using anholonomic bases and describing geometry in terms of arbitrary tangent space frames. For example, the cosmological perturbations of teleparallel models \cite{GK}  can be equivalently described in the covariant language \cite{covpert}. Moreover, using a non-vanishing spin connection might sometimes be very convenient \cite{issues}, because of allowing us to use a much simpler tetrad.

Given a defining covariantly-constant tetrad field $e^a_{\mu}(x)$, we can express any other tetrad $h^a_{\mu}$ by a {\it local} general linear transformation:
$$h^a_{\mu}(x)= L^a_b (x) \cdot e^b_{\mu}(x).$$
Then we get 
$$\bigtriangledown_{\mu} h^a_{\nu}=(\partial_{\mu} L^a_b) \cdot e^b_{\nu}=(\partial_{\mu} L^a_b) (L^{-1})^b_c h^c_{\mu}.$$
In case of $L^a_b$ being a matrix from the Lorentz group,  in the r.h.s. we immediately recognise our usual flat spin-connection, and the so called "tetrad postulate" in the equation as a whole. Moreover, the spin-connection would be precisely the one to be used for differentiating the vectors in an anholomic frame $A^a=A^{\mu}h^a_{\mu}$.

Of course, we can define the teleparallel geometry in terms of an arbitrary tetrad $h^a_{\mu}$ and a spin-connection which is associated to it in a Lorentz-covariant way. No problem with that. Neither metric nor torsion, either in the equations of motion or in the action, depend on which  approach is chosen; and one can even make an algebraic change of variables absorbing the Lorentz matrices into the tetrad which turns the model into its  cousin without spin-connection \cite{BFGG}. However, it would be much more beautiful to simply start by directly indicating which tetrad $e^a_{\mu}$ has it vanishing. Moreover, this is precisely the meaning of teleparallelism: there exists a  basis of parallelly transported vectors. Analogously, it is absolutely all right, and often quite useful, to deal with Euclidean spaces in curvilinear coordinates, however it is also important to understand that the Cartesian coordinates are an important defining feature of these spaces.

To state it once more, I don't claim that there is something wrong with the covariant approach. It is fully equivalent to the pure tetrad one. However, it is not necessary either. The pure tetrad description is better suited to the geometrical meaning of the Weitzenb{\" o}ck connection. It uses precisely the tetrad which represents, up to global transformations, the notion of teleparallelism. At the same time, it is of course possible to rewrite the whole story in terms of an arbitrary basis in the tangent space, thus getting the covariant formulation; possible but by no means necessary.

\section{Conclusions}

Nowadays we have a plenty of modified gravity models, maybe even too many of them \cite{fate}. But even if way too many indeed, they are not to simply be neglected. Some unexpected models might finally be good for phenomenology, and a serious look at all of them does really help us understand the mathematical properties of gravitational theories better. And it is even more so about the models which employ modifications of the very geometric foundations of gravity, such as modified teleparallel frameworks.

In my opinion, all the simplest modified teleparallel models are highly problematic and might be just unviable. However, it is still very important to understand how and why they work or don't work. In particular, one of the very hot topics in the recent years has been their local Lorentz symmetry violation. My claim is that, in a well-defined meaning, it is violated for sure. But as always, the symmetry can be formally restored by a kind of St{\"u}ckelberg trick. In this paper I argue that this restoration, even though it might often be useful and nice, is very artificial in its nature, and therefore should not be taken as a must, unless one day it gets required by some other tasks. 

Still, there are many other unresolved issues of modified teleparallel theories. For example, even leaving alone the foundational issues, one could say that we need a spin-connection for coupling the fermions. However, for that we would anyway need to consider the one corresponding to the Levi-Civita connection of the metric which is possible but, I would say, inelegant in the teleparallel realms. Maybe, with fermions or without, some more profound theory will justify necessity of the covariant approach. However, it is not the case for the currently studied modifications of teleparallel gravity.

{\bf Acknowledgment.} It is very important to discuss the basics, as well as achievements and troubles of many new modified gravity models. I am very grateful to the organisers of the Metric-Affine Frameworks for Gravity 2022 conference in Tartu, for it was a very nice and useful opportunity to share the viewpoints and have vivid discussions with many great colleagues.

\end{document}